\documentclass[11pt]{article}
\usepackage{moriond,epsfig}

\def\Journal#1#2#3#4{{#1} {\bf #2}, #3 (#4)}

\def\MNRAS{\em MNRAS}
\def\apj{\em ApJ}
\def\apjs{\em ApJS}
\def\aj{\em AJ}
\def\RMP{\em  Rev. Mod. Phys}

\def\PRD{{\em Phys. Rev.} D}

\def\be{\begin{equation}}
\def\ee{\end{equation}}
\def\bea{\begin{eqnarray}}
\def\eea{\end{eqnarray}}

\begin{document}
\vspace*{4cm}
\title{$\Lambda$CDM and the WMAP power spectrum beam profile sensitivity}

\author{ Utane Sawangwit \& Tom Shanks  }

\address{Department of Physics, Durham University, South Road,\\
Durham DH1 3LE, England}

\maketitle\abstracts{We first discuss the sensitivity of the WMAP CMB power
spectrum to systematic errors by calculating the raw CMB power spectrum from
WMAP data. We find that the power spectrum is surprisingly sensitive to the WMAP
radiometer beam profile even at the position of the first acoustic peak on
$\approx1$ degree scales. Although the WMAP beam profile core is only $12.'6$
FWHM at W, there is a long power-law tail to the beam due to side-lobes and this
causes significant effects even at the first peak position. We then test the
form of the beam-profile used by the WMAP team which is based on observations of
Jupiter. We stacked  radio source beam profiles as observed in each WMAP band
and found that they showed a wider profile in Q, V, W than the Jupiter profile.
We have now checked that this is not due to any Eddington or other bias in our sample by
showing that the same results are obtained when radio sources are selected at
1.4GHz and that our methods retrieve the Jupiter beam when it is employed in
simulations. Finally, we show that the uncertainty in the WMAP beam profile
allows the position as well as the amplitude of the first peak to be  changed and how
this could allow simpler cosmologies than standard $\Lambda$CDM  to fit the CMB
data. }

\section{Introduction}
The  standard $\Lambda$CDM cosmological model is a quite perplexing mixture of
impressive observational successes (e.g. Hinshaw {\it et al}.~\cite{gh03},
Komatsu {\it et al}.~\cite{ko}, Hicken {\it et al}.~\cite{hick}, Kessler {\it et
al}.~\cite{kess}) coupled with wider implications which make the model
complicated to the point of implausibility (eg Weinberg~\cite{sw}). Some
fundamental and astrophysical  issues for the standard model are as follows:

\begin{itemize}
\item The exotic, weakly interacting, particles envisaged as candidates for the dark matter component of the 
standard model are  still undetected in the laboratory (eg Aprile {\it et al}.~\cite{ap}).
\item The inclusion of a cosmological constant means that ratio of the 
vacuum energy density to the radiation energy density after inflation is 1 part  in $10^{100}$, a 
fine-tuning coincidence which leads to appeals to the anthropic principle for an explanation (eg Efstathiou~\cite{gpe}).
\item Even if fine-tuning arguments are regarded as unsatisfactory, the problem is that 
inflation was set up to get rid of fine-tuning in terms of the `flatness' problem (Guth~\cite{guth}) and so the introduction of more fine-tuning 
with the cosmological constant appears circular.
\item $\Lambda$ has the wrong sign for string theorists who prefer a negative $\Lambda$ than a 
positive $\Lambda$, ie a cosmology which is approximately Anti-de Sitter rather than de Sitter (eg Witten~\cite{ew}).
\item The standard inflationary  model predicts not just one but $10^{{10}^{77}}$ Universes (Steinhardt~\cite{pjs}).
\item Astrophysically, any CDM model in the first instance predicts a featureless mass function for galaxies 
whereas the galaxy luminosity  function shows a sharp `knee' feature (eg Benson {\it et al}.~\cite{ajb}).
\item CDM models predict that large structures should form last and therefore should be young 
whereas, observationally, the largest galaxies and clusters appear old (eg Cowie {\it et al}.~\cite{llc}).
\item To fix the above two problems, large amounts of feedback (eg Bower {\it et
al}.~\cite{rgb}) are invoked which results in more energy now being used to
prevent stars forming than in forming them under gravity.
\end{itemize}

\section{WMAP CMB Power Spectrum}\label{subsec:prod}

The above  issues  mean that the standard model requires remarkable observations
in its  support and the most remarkable of these is represented by the acoustic
peaks in the CMB power-spectrum as measured by WMAP (eg Hinshaw {\it et al}.~\cite{gh09})
 and other CMB experiments.
Much therefore depends on the accuracy of these observations and in particular
on the position of the first acoustic peak at wavenumber l=220 or $\approx1$deg.
A first peak at scales as large as these strongly favours a CDM model. Attempts
have been made to move the first peak using cosmic foregrounds such as large
clusters via the SZ effect (Myers {\it et al}.~\cite{adm}, Bielby {\it et al}.~\cite{rmb}) 
or gravitational lensing (Shanks~\cite{ts}) or even inhomogeneous
reionisation at $z\approx10$ but generally the effects have been small.

\begin{figure}
\centering
\psfig{figure=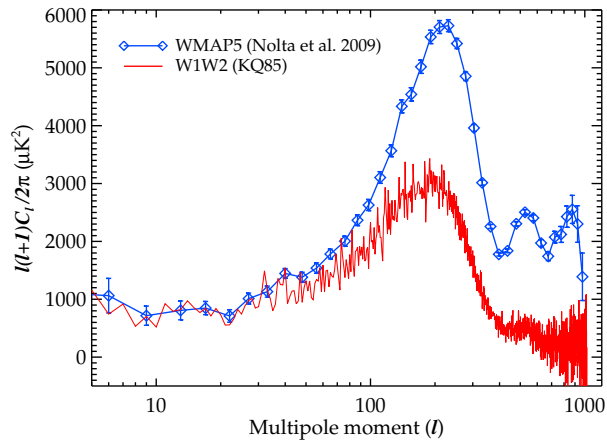,height=2.5in}
\caption{The red line shows the raw WMAP W band power spectrum estimated from
the cross-correlation of the WMAP5 W1 and W2 maps. The blue diamonds $+$ line shows the 
final WMAP5 spectrum after
`de-beaming' using the Jupiter beam( $+$`cut-sky' correction). The large effect of de-beaming even at the
first acoustic peak ($\approx1$deg) is caused by beam sidelobes, even though the
beam's Gaussian core has a width of only $12.'6$ FWHM.
\label{fig:one}}
\end{figure}

At first sight, it seems unlikely that  any observational effect of the
resolution of CMB radiometers such as those used by WMAP on the position and
amplitude of the first peak could be significant. The highest resolution of the
WMAP satellite comes at the 94GHz W band where the core of the beam profile has
$12.'6$ FWHM (eg Page {\it et al}.~\cite{lp}) and it seems unlikely that such a
narrow beam profile would have an effect at the $\approx1$ degree scale of the
first peak. However, Sawangwit and Shanks~\cite{ss} have recalculated the WMAP5
power spectrum, $C_l$,  by cross-correlating the maps from the W1 and W2
detectors as an example. This raw spectrum is compared in Fig. 1 to the WMAP5
spectrum that is usually fitted by the standard model and large differences can
be seen even at the scale of the first acoustic peak, where the raw spectrum is
a factor of $\approx2$ lower than expected. Most of the reason for this
difference ($\approx70$\%) is the effect of the WMAP beam profile with the
remainder being due to sky incompleteness. Although the W band beam profile has
its narrow, $12.'6$, core, it also has wide sidelobes which fall off as a power-law
with angle, rather than as a Gaussian. The WMAP beam profile is usually
estimated by measuring bright planetary sources such as Jupiter and the beam
must be known to high accuracy at 1 degree  scales where the profile reaches
0.1\% of its peak value. Thus the effects of `de-beaming' are $\approx70$\% of
the first peak height and even more at the position of the second and third
peaks.

\section{Testing the WMAP beam using radio sources}

The sensitivity of the WMAP power spectrum to the beam profile suggests that it
is important to test the profile used by the WMAP team. Sawangwit and Shanks~\cite{ss}
stacked the CMB data at the positions of $\approx150-250$ WMAP5 radio sources as
catalogued by Wright {\it et al}.~\cite{nedw}, excluding all sources identified as extended 
in ground-based, higher resolution 5GHz surveys (see Fig. 2). They found that at
the Q, V and W bands, the stacked source profiles looked increasingly broad and
broader than the Jupiter profile on scales of $10-30'$. Beyond these scales, the
noise on the stacked radio-source profile means that little information about
the profile can be obtained by this method. Sawangwit \& Shanks~\cite{ss} also found more
marginal evidence that the profile width may increase as  source flux decreases,
possibly suggesting that there was a non-linearity in the WMAP flux scale.

\begin{figure}
\centering
\psfig{figure=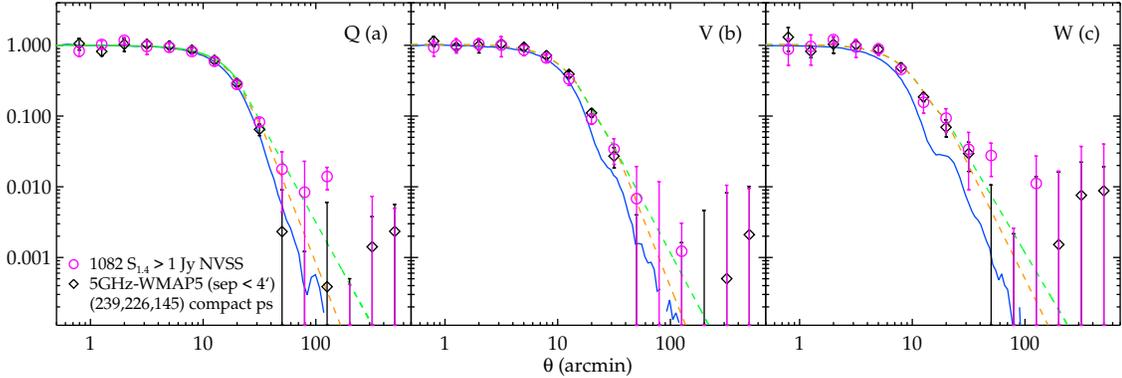,height=2.2in}
\caption{Stacked radio source beam profiles for the Q, V and W WMAP bands. Open
black diamonds represent sources from the WMAP catalogue of Wright {\it et al}.
and open magenta circles represent NVSS sources. The Jupiter beam is shown in
each case (blue solid line). The radio source profiles appear wider than the
Jupiter profiles particularly in the V and W bands. The dashed green and red
lines are the power-law fits of Sawangwit and Shanks to the observed WMAP radio
source profiles.
\label{fig:two}}
\end{figure}

In Fig. 2 we also show  a new WMAP stack on the positions of a flux limited
selection of NVSS sources at the 1.4GHz frequency, now rejecting any source with
a neighbour within a 1 degree radius. As can be seen,  we find an almost
exactly similar result to that based on selecting sources from the WMAP5
catalogue. We have further found that the result reproduces in the sample of
sources detected using the `CMB-free' method of Chen and Wright~\cite{chen}.
Sawangwit~\cite{us} has also checked our result in 100 CMB sky simulations, each
containing a similar  number of sources as in the WMAP5 source list and with
each source assuming the Jupiter beam profile. We found that in this simulated
case our stacked radio source profile retrieved the Jupiter beam almost
perfectly. All of these tests suggest that our results are not subject to any
`Eddington bias' or any other  bias.

\section{Other models that fit the WMAP CMB Power Spectra}

Sawangwit and Shanks~\cite{ss} fitted power-laws to the WMAP radio source beam
profile (see Fig. 2) and showed how sensitive the height of the first acoustic
peak is to relatively small deviations away from the Jupiter profile. Here we
show that  beam profiles  that are consistent with the radio sources can also
significantly shift the {\it position} of the first peak from the standard
$l=220$ multipole to  $l>300$. We do this by reverse engineering a beam profile,
$b_s(\theta)$, from the square of the beam window function, $b_l$, obtained by
dividing the $C_l$ of a model with $\Omega_{baryon}=1$,
$H_0=35$km~s$^{-1}$Mpc$^{-1}$ and $\tau=0.35$ (Shanks~{\cite{ts2,ts3,ts4}) by
the `raw', W1W2 $C_l$ shown in Fig. 1. Transforming back to the beam profile,
$b_s(\theta)$, leaves a profile which shows a `ringing' at large $\theta$ and
which oscillates between positive and negative values. After only taking the
small-scale  part of the profile which is positive and `squeezing' the profile
to smaller scales by 25\% to compensate for the loss of the negative parts gives
us a `do-it-yourself' profile for the large scales. At $\theta<20'$ we fit the
radio source profile with a Gaussian and then an exponential which helps match
the Gaussian  smoothly onto the large scale part. The resulting profile is shown
in Fig. 3, where it is compared to the WMAP and NVSS  radio source W band
profile and the Jupiter profile. It is the spike at $\approx35'$ which is vital
to move  the $l=220$ peak to match the $\Omega_{baryon}=1$, $C_l$ peak at
$l=330$. The resulting `diy' debeamed W band $C_l$ compares well to the
theoretical $\Omega_{baryon}=1$, low $H_0$ $C_l$  on which it is based (see Fig.
4); the raw W1W2 $C_l$ is also shown. Beyond $l=600$ where the S/N for the WMAP
spectrum drops, the  $C_l$ from the QUAD experiment (Brown {\it et
al}.~\cite{mb}) is also plotted. Again, we see broad agreement with the
$\Omega_{baryon}=1$ model, although there is some detailed disagreement with
QUAD peak positions at $l>1000$.

 \begin{figure}[hbp] Ê 
\begin{minipage}[b]{0.5\linewidth} 
\centering \psfig{file=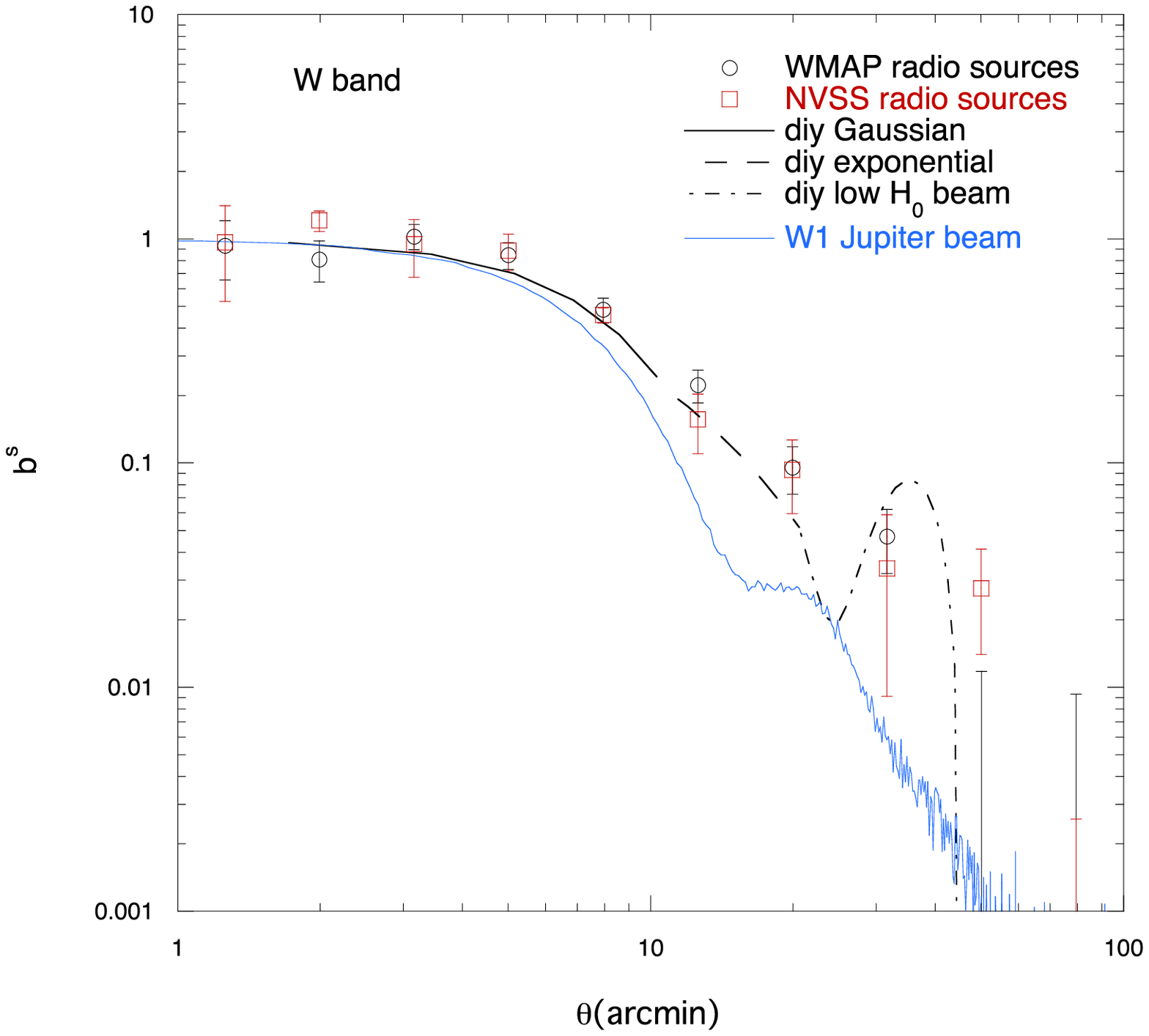,height=2.5in} Ê 
\caption{A partly `reverse-engineered' or `do-it-yourself' WMAP W band beam profile 
comprising a Gaussian in the centre (solid black line), then an exponential (dashed black line)
and then a spike (dot-dashed black line). This beam is not inconsistent with 
the radio source profiles and produces the $C_l$ given by the red solid line in Fig. 4. } Ê 
\end{minipage} Ê 
\begin{minipage}[b]{0.5\linewidth} Ê 
\centering \psfig{file=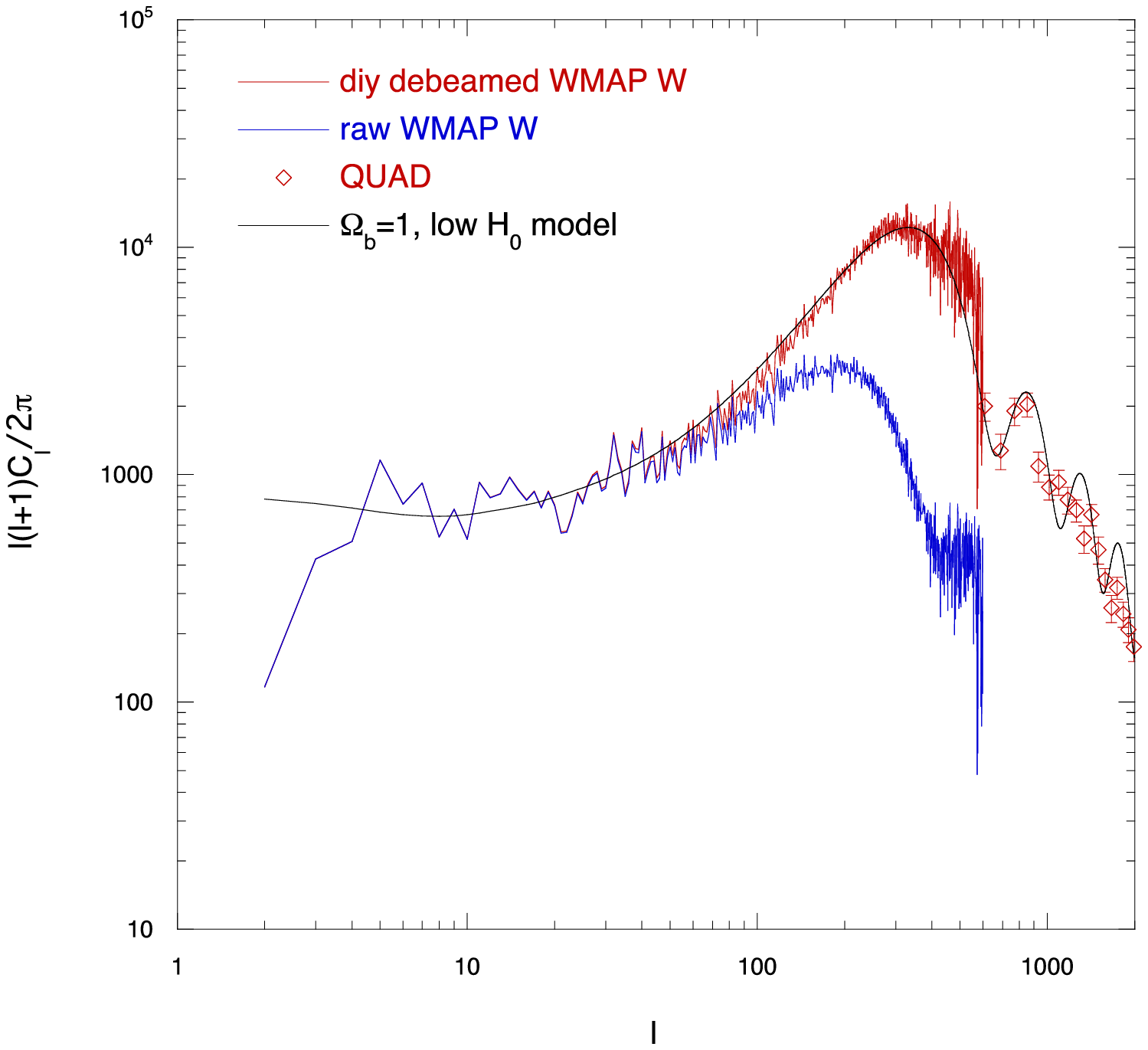,height=2.5in} Ê 
\caption{ The $C_l$ debeamed using the `diy' beam from Fig. 3  is shown as the
red line and the $\Omega_{baryon}=1$, low $H_0$ model  and the raw WMAP5 W1W2
$C_l$ from which it was partly reverse engineered are shown as the solid black
and blue lines. This model is also compared to the QUAD results at $l>600$ (red
open circles). } Ê
\end{minipage}Ê 
\end{figure}

\section{Conclusions}

We have discussed  various fundamental and astrophysical problems for the
standard $\Lambda$CDM cosmological model, including its requirement for
undiscovered physics in terms of the  weakly interacting CDM particle and in the
one part in $10^{100}$ fine-tuning in terms of the small size of the
cosmological constant. But the model has received overwhelming support from CMB
experiments such as WMAP in terms of the position of the first acoustic peak in
the power spectrum. However, Sawangwit and Shanks~\cite{ss} have shown the high
sensitivity of the amplitude of the first acoustic peak to the detailed form of
the WMAP instrumental beam profile and that stacking unresolved  radio sources
indicates wider beams than expected in the Q,V and W bands. Here we have shown
that a WMAP W band beam which is not inconsistent with the radio source beam profile
can also reproduce the power spectrum of a simple inflationary model with
$\Omega_{baryon}=1$ and a low value of $H_0$. This model produces a peak at
multipole $l=330$ rather than $l=220$. It also reproduces the broad
form, although not the detailed peak positions, of the observed power spectrum
in the $600<l<2000$ region from e.g the  QUAD experiment. However, the $l=330$
WMAP peak was basically `reverse-engineered' to fit the $\Omega_{baryon}=1$
model, using the freedom afforded by the loose constraints on the stacked  radio
source profiles at scales $\theta>30'$ and it remains to be seen whether the
actual WMAP beam  is consistent with this very simple cosmological model.

\section*{Acknowledgments}
We acknowledge the WMAP team for making their data freely and publicly available. 
Utane Sawangwit acknowledges a Royal Thai Government Scholarship for  PhD funding.

\end{document}